\documentclass[seceq]{ptptex}

\usepackage{graphicx}
\usepackage{amsfonts}
\usepackage{amsmath}
\usepackage{bbm}

\newcommand{\rmO}{{\mathcal{O}}}

\newcommand{\eq}{eq.~}
\newcommand{\eqs}{eqs.~}

\renewcommand{\vec}[1]{{\bf #1}}
\newcommand{\nwc}{\newcommand}

\nwc{\nl}  {\newline}
\nwc{\be}  {\begin{equation}}
\nwc{\ee}  {\end{equation}}
\nwc{\bmu} {\bar{\mu}}
\nwc{\ba}  {\begin{eqnarray}}
\nwc{\ea}  {\end{eqnarray}}
\nwc{\bc}  {\begin{center}}
\nwc{\ec}  {\end{center}}
\nwc{\bi}  {\begin{itemize}}
\nwc{\ei}  {\end{itemize}}
\nwc{\nn}  {\nonumber\\}
\nwc{\Tr}  {\mathop{\rm Tr}}
\nwc{\re}  {\mathop{\rm Re}}
\nwc{\im}  {\mathop{\rm Im}}
\nwc{\Hc}  {\mathop{\rm H.c.}}
\nwc{\la}[1]{\label{#1}}
\nwc{\rmi}[1]{{\mbox{\scriptsize #1}}}
\nwc{\nr}[1]{(\ref{#1})}
\nwc{\fr}[2]{{\frac{#1}{#2}}}
\nwc{\msbar}{\overline{\mbox{\rm MS}}}
\nwc{\lambdamsbar}{\Lambda_{\overline{\rm MS}}}
\nwc{\dr}{{4d\to3d}}

\newcommand{\Nc}{N_{\rm c}}

\newcommand{\rmii}[1]{{\mbox{\tiny\rm{#1}}}}

\def\lsi{\raise0.3ex\hbox{$<$\kern-0.75em\raise-1.1ex\hbox{$\sim$}}}
\def\gsi{\raise0.3ex\hbox{$>$\kern-0.75em\raise-1.1ex\hbox{$\sim$}}}
\nwc{\lsim}{\mathop{\lsi}}
\nwc{\gsim}{\mathop{\gsi}}

\def\NN{{\rm I\kern -.16em N}}
\def\II{{\rm I\kern -.16em I}}
\def\RR{{\rm I\kern -.2em  R}}
\def\ZZ{Z \kern -.43em Z}
\def\QQ{{\rm \kern .25em
             \vrule height1.4ex depth-.12ex width.06em\kern-.31em Q}}
\def\CC{{\rm \kern .25em
             \vrule height1.4ex depth-.12ex width.06em\kern-.31em C}}

\newcommand{\zz}{{\mathbb{Z}}}

\title{
On bulk viscosity and moduli decay%
}

\author{
Mikko \textsc{Laine}%
}

\inst{
Faculty of Physics, University of Bielefeld, 
D-33501 Bielefeld, Germany
}

\abst{
This pedagogically intended lecture, 
one of four under the header ``Basics of thermal QCD'', 
reviews an interesting relationship, 
originally pointed out by B\"odeker, that exists between 
the bulk viscosity of Yang-Mills theory (of possible 
relevance to the hydrodynamics of heavy ion collision experiments) 
and the decay rate of scalar fields coupled very weakly to a heat bath
(appearing in some particle physics inspired cosmological scenarios).
This topic serves, furthermore,  as a platform on which 
a number of generic thermal field theory concepts are illustrated.  
The other three lectures (on the QCD equation of state and the rates
of elastic as well as inelastic processes experienced by heavy quarks) 
are recapitulated in brief encyclopedic form.  
}

\begin{document}

\maketitle

%
%

%
\section{Introduction and outline}

It has been a longstanding dream that experimental tests of thermal QCD
through heavy ion collision experiments could yield theoretical insights
that would be useful for some cosmological problems as well. These
lectures covered selected topics within thermal QCD with this perspective
in mind. The observables touched upon were the equation of state, shear
and bulk viscosities, as well as the rates of elastic and inelastic 
reactions experienced by heavy quarks. Depending on the observable 
the focus was either on elaborating on the basic concepts, on outlining 
the link between heavy ion collisions and cosmology, or on reviewing 
modern developments.

Because of severe page limitations, it is not possible to cover 
all of the lectures in a detailed from in the proceedings. Rather, 
the choice has been made to concentrate on the second lecture, 
which is presented in some detail. The second lecture has been chosen
because, first of all, the topic is rather elegant and concise, yet 
apparently little-known. Furthermore, 
it offers an opportunity to present 
a result which to my knowledge is original, 
generalizing the analysis of ref.~\cite{db}
from a scalar field theory to a Yang-Mills theory. Finally, 
this topic can serve as a simple example 
permitting to introduce several tools 
(Kubo relations, Euclidean and Minkowskian correlators, the concept
of transport coefficients) that are applicable much more widely 
in thermal field theory than to the very problem at hand. 

The presentation is organized as follows. 
In section~\ref{se:lec1} we review the current theoretical understanding
concerning the equation of state of QCD and the role that it 
may play in heavy ion collisions and in cosmology. 
Section~\ref{se:lec2}, the main part of this presentation, 
concentrates on the topic announced in the title. 
The subject of section~\ref{se:lec3} is the rate of kinetic thermalization
of heavy quarks, caused by elastic scatterings with the light
degrees of freedom in the hot plasma. Finally, in section~\ref{se:lec4},
recent developments in thermal heavy quarkonium physics, which to 
some extent can also be associated with the rate of inelastic 
processes felt by heavy quarks, are reviewed.

%
\section{Lecture 1: QCD equation of state}
\la{se:lec1}

In cosmology, it is in general 
an excellent approximation to set the baryon chemical potential to zero, 
so that thermodynamic potentials are functions of the temperature, 
$T$, only. Then the expansion rate of the system and the production 
rates of weakly interacting particles from the thermal medium
(such as Dark Matter) depend on $p(T), p'(T)$ and $p''(T)$, 
where $p(T)$ denotes the pressure~\cite{gg}. The same
functions (save without the contribution of photons and leptons, which have no 
time to thermalize) dictate the expansion of the thermal fireball 
created in an energetic heavy ion collision. Hence the longstanding
efforts, both analytic (via chiral perturbation theory at low 
temperatures, $T \ll 200$~MeV, and the weak-coupling expansion 
at high temperatures, $T \gg 200$~MeV) and numerical (via lattice
Monte Carlo simulations), to reliably determine these functions. 
At temperatures below one GeV, relevant for 
heavy ion collisions, the major qualitative finding, 
all but dashing
early hopes of a spectacular scenario~\cite{ew}, 
has been that there is probably {\em no phase transition} 
at any~$T$, only a smooth crossover.\cite{co} Nevertheless large-scale
numerical efforts to determine $p(T)$ and its derivatives for physical
QCD in the infinite-volume and continuum limit go on.
At temperatures above a few GeV, relevant for cosmology
(only exceptional scenarios operate below this temperature range~\cite{brs}), 
the system should be addressable with weak-coupling techniques. 
Perturbation theory does suffer from serious infrared problems, 
however: besides proceeding in powers of 
$ 
 \alpha_s^{\scriptscriptstyle 1/2}
$
rather than $\alpha_s$,\cite{jk} it also comes with 
non-perturbative coefficients starting at the order $\alpha_s^3$.\cite{ir} 
That some coefficients are non-perturbative does not mean that 
they are beyond reach: indeed the non-perturbative input~\cite{plaq} 
can be determined via the use of effective field theory 
techniques~\cite{dr}. As of today the non-perturbative part
of the $\alpha_s^3$-term is known,\cite{lattg6} and 
it is possible to compile (systematically improvable) phenomenological 
values for $p(T),p'(T)$ and $p''(T)$ of the Standard Model, 
applicable in the whole temperature range from $\sim$ GeV up to 
a few hundred GeV\cite{pheneos}, or even higher if necessary.\cite{gv}

%
\section{Lecture 2: bulk and shear viscosities}
\la{se:lec2}

As a second example on the interplay between heavy ion collision
experiments, cosmology, weak-coupling thermal field theory, and 
non-perturbative lattice simulations, we discuss observables 
known as the bulk and shear viscosities. 

%
\subsection{Phenomenological background}

The bulk and shear
viscosities are, in general, functions of the temperature and 
chemical potentials and parameterize {\em gradient corrections}
to the energy-momentum tensor of a multiparticle system close to 
thermal equilibrium. At zeroth order in the gradient expansion 
one considers a system in which the temperature, $T$, and 
the four-velocity of the (``one-component'') medium, $u^\mu$, 
are constant in temporal and spatial coordinates; then 
the energy-momentum tensor, $T^{\mu\nu}$, is constructed
out of symmetric and covariant structures proportional to 
the metric tensor, $g^{\mu\nu}$, and to $u^{\mu}u^{\nu}$. 
At first order in gradients, the structure 
$\partial^{\mu} u^\nu + \partial^\nu u^\mu$
can appear as well. Thus, $T^{\mu\nu}$ has 
the form (see, e.g.,\ ref.~\cite{ll})
\be
 T^{\mu\nu} = (p+e) u^\mu u^\nu - p\, g^{\mu\nu} + 
 \rmO((\eta,\zeta) (\partial^\mu u^\nu + \partial^\nu u^\mu ))
 \;. \la{Tmunu}
\ee
Here $p$ denotes the pressure, $e$ the energy density, 
$\eta$ the ``shear'' viscosity, and $\zeta$ the ``bulk'' viscosity,
and the metric convention ($+$$-$$-$$-$) was assumed. The equation
of motion reads $\partial_\mu T^{\mu\nu} = 0$. 

More precisely, 
the shear viscosity coefficient $\eta$ is defined
to be a function that multiplies the traceless part 
of the leading gradient correction; the bulk viscosity 
coefficient $\zeta$ multiplies the trace part.  
The explicit forms of the corresponding 
structures are most simply displayed in 
a non-relativistic frame, $|u^i| \ll 1$; then 
\be
 T_{ij} \approx (p - \zeta \nabla\cdot\vec{v})\delta_{ij} 
 - \eta (\partial_i v^j + \partial_j v^i 
       - \fr23 \delta_{ij} \nabla\cdot\vec{v} )
 \;, 
\ee
where $\nabla\cdot\vec{v} = \partial_i v^i$ and $\delta_{ij}$
is the usual Kronecker symbol.

Now, since the system generated in a heavy ion 
collision is relatively {\em small}, gradients can be large, 
with a scale given by the system size: 
\be
 \left( \frac{\partial u}{u} \right)^{-1} 
 \sim  {\mbox{few~fm}} \sim \frac{\mbox{few}}{T_\rmii{QCD}}
 \;, \la{scale_hic}
\ee 
where $T_\rmii{QCD} \equiv 200$~MeV.
In the relativistic
limit, $\eta\sim T^3$ and $p\sim T^4$; this then implies 
that $\eta \, \partial u  \sim p\, u^2$. In other words, gradient
corrections could be as large as the leading terms. Setting
aside the question of whether a gradient expansion can be
justified at all in this situation, it seems conceivable
that the viscosities could indeed affect the 
expansion of the thermal fireball created in a heavy ion 
collision. An example on the effects of the shear viscosity
on the azimuthal anisotropies of particle spectra produced
in a heavy ion collision can be found in ref.~\cite{rr}  
while possible implications of a (large) bulk viscosity
have been explored in, e.g.,\ ref.~\cite{kr}.

We now turn to cosmology. An immediate qualitative difference 
between the matter created in a heavy ion collision and that 
filling the Early Universe is that the latter system is rather 
homogeneous. For instance, if we are interested in the {\em overall 
expansion}, then the system size can be identified with 
the horizon radius, which in the QCD epoch ($T\sim T_\rmii{QCD}$) is 
\be
 \ell_H  \sim t
 \sim \frac{m_\rmi{Planck}}{T^2} 
 \sim \frac{10^{20}}{T_\rmii{QCD}}
 \;. \la{scale_cos}
\ee
This can be contrasted with \eq\nr{scale_hic};
thus gradient corrections are bound 
to be insignificant as far as the {overall expansion} is
concerned. (Nevertheless, there may be other 
physics questions for which they do play a role; 
indeed some inhomogeneities do 
exist within the horizon, as is famously indicated by 
anisotropies in the cosmic microwave background and as is required for
successful large-scale structure formation. 
In spite of the fact that at early times
their relative magnitude is of order $\sim 10^{-5}$ rather than of 
order unity, viscosities might still affect the 
evolution of density perturbations;   
see, e.g.,\ refs.~\cite{sw,hss}.)

It turns out, however, that the bulk viscosity $\zeta$ makes 
a {\em formal} appearance in a completely different context.\cite{db} 
In order to discuss this, we need to next recall some basic facts
concerning the important role that scalar fields play in cosmology. 

%
\subsection{Scalar fields and the moduli problem in cosmology}

Although no fundamental scalar fields have been experimentally 
discovered up to date, they play a very important role in inflationary
cosmology, and are also considered to be a generic feature of many models
inspired by string theory or supersymmetry. 
At an early stage, it is assumed that 
a significant amount of energy may be 
stored in the scalar fields. Later on, 
as the Universe enters the radiation dominated epoch, the scalar
fields should ``decay'', i.e.\ transform their energy to the known
particles (electrons, photons, etc);  this is called a reheating period. 
It is possible to imagine scalar fields, however, which 
are coupled so weakly that they decay later than the inflaton field, 
or not at all. In the case of a delayed decay, one may use
the scalar field to generate a possibly observable ``second'' 
component of density perturbations; such a scalar field is 
often called a ``curvaton''.  
Yet if the decay is too slow and 
the scalar field does not get rid of its energy density at all, 
it eventually becomes a dominant component, in conflict with 
observation. Such a situation is met particularly 
in connection with scalar fields called {\em moduli}, and 
the problem of their slow decay is the ``moduli problem''.

We note, in passing, that there 
are many other similar ``dangerous relics'' in cosmology: 
in particular theories predicting the generation of topological
defects in the form of domain walls or monopoles are practically 
excluded, if the associated energy scale corresponds to physics
beyond the Standard Model.

To be concrete, let us consider a toy model in which 
a scalar field, $\varphi$, couples to Yang-Mills fields: 
\be
 \mathcal{L} =  \fr12 \varphi (-\square - m^2 ) \varphi
 - \frac{1}{4 g^2} F^{a\mu\nu}F^a_{\mu\nu}
 - \frac{\varphi}{M} \times F^{a\mu\nu}F^a_{\mu\nu}
 \;, \la{L}
\ee
where $m$ is some ``small'' mass scale, 
perhaps $m \sim m_\rmi{SUSY}$, while $M$ is some 
large scale, perhaps $M\sim m_\rmi{Planck}$. The non-renormalizable
coupling between $\varphi$ and the Yang-Mills fields is suppressed
by the heavy mass scale; hence the coupling between the two sets
of fields is very weak. In ref.~\cite{db} a similar system was 
considered, however the scalar field $\varphi$ was coupled to another
scalar field, denoted by $\chi$.

If we now assume that the initial state of the system 
is such that $\langle \varphi(0) \rangle \gg m$, then the 
initial energy density in the scalar field 
is $e_\varphi \gg m^4$. According to \eq\nr{L}
the scalar field can decay to gauge bosons; however the coupling
is suppressed by $M$, so the (vacuum) 
decay rate is $\Gamma \sim m^3 / M^2$.
The corresponding time scale, $t_\varphi \sim M^2/m^3$, 
can be contrasted with that in \eq\nr{scale_cos}; it is known from 
observation that a conventional radiation-dominated expansion is needed 
at least starting from the nucleosynthesis epoch 
($T\sim T_\rmi{n} \equiv 0.1$~MeV).
(For completeness, a cartoon of a standard cosmological 
scenario is attached as fig.~\ref{fig:history}.)
If $M\sim m_\rmi{Planck}$, the corresponding 
constraint $t_\varphi \ll t_\rmi{n}$ translates 
to $m \gg (m_\rmi{Planck} T_\rmi{n}^2)^{1/3} \sim 10$~TeV. 
While not impossible, this looks unattractive if $m$ is to 
be related, for instance, 
to some lifted SUSY flat direction.  

\begin{figure}[tb]


\centerline{\includegraphics[height=8.0cm]{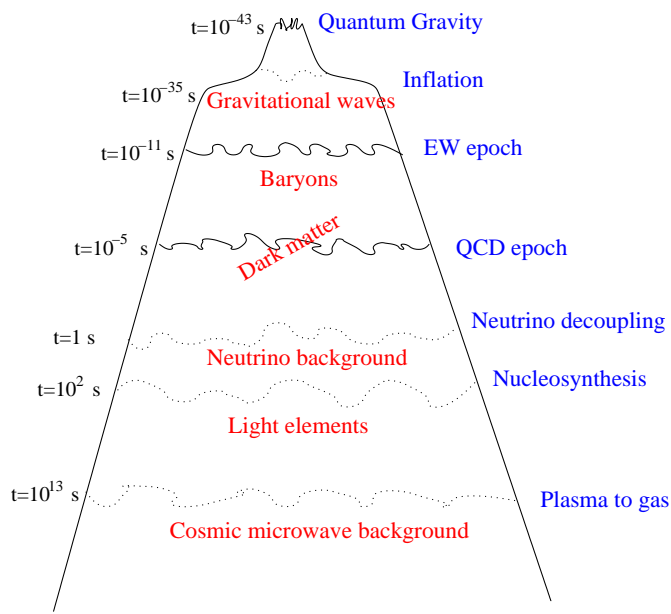}}

\caption{A cartoon of a possible cosmological scenario.
The age of the Universe, $t$, and the temperature of 
the plasma filling it, $T$, are related
through $t[\mbox{s}]\sim (T[\mbox{MeV}])^{-2}$
(cf.\ \eq\nr{scale_cos}).}


\la{fig:history}
\end{figure}

%
\subsection{Decay rate in thermal field theory}

The question now is, could the 
vacuum decay rate ($\Gamma \sim m^3 / M^2$) be modified 
by thermal corrections associated with the ``normal'' degrees 
of freedom, represented by $F^a_{\mu\nu}$? In particular, could 
the thermal plasma be ``viscous'' enough to dissipate the energy
related to the oscillations of the  ``dangerous relic'' $\varphi$ 
well before the nucleosynthesis epoch, rendering $\varphi$ harmless
even if the mass $m$ were only in the TeV range or below it? 

In order to answer this question, let us consider the 
equation of motion satisfied by $\varphi$. In the temperature 
range $m \ll T \ll M$, the scalar field is ``light'' and oscillates
slowly compared with the typical frequencies of the thermal
degrees of freedom. In this situation, we can assume that the 
``fast'' thermal fields see $\varphi$ essentially as a constant, 
and have time to adjust to any given value. From their perspective, 
then, the system remains in thermal equilibrium, only in the presence
of a background field $\varphi$. From the perspective of $\varphi$, 
on the other hand, the net effect of the thermal medium is to exert
some friction which slows down its oscillations. Focussing on the 
infrared modes of $\varphi$ (the modes with the slowest temporal 
and spatial variations), we can expect an equation of motion of the form
\be
 \square \varphi + V_\rmi{eff}'(\varphi) = - \Gamma \dot{\varphi}
 + \rmO(\raise-0.15em\hbox{$\stackrel{\dots}{\varphi}$}, 
 \nabla^2\dot\varphi,
 \dot{\varphi}^2,(\nabla\varphi)^2)
 \;. \la{eom}
\ee
The main task then is to determine the coefficient $\Gamma$
(although we will also discuss thermal corrections to 
$V_\rmi{eff}$, particularly through a mass parameter
$m_\rmi{eff}^2$).

Now, a standard tool in thermal field theory is a ``Kubo formula'', 
which indeed allows to determine ``response'' or ``transport'' 
coefficients of the same type as $\Gamma$. Note that $\Gamma$
as defined by \eq\nr{eom} is, by construction, a {\em constant}.
It is given by   
\be
 \Gamma = \lim_{\omega\to 0} \frac{\rho(\omega,\vec{0})}{\omega} 
 \;, \la{Kubo}
\ee
where, for $\mathcal{L}_\rmi{int} =  - \varphi H_\rmi{int}$, 
$\rho$ is a {\em spectral function} related to the operator 
$\hat H_\rmi{int}$: 
\be
 \rho(\omega,\vec{0})
 \equiv \int_{t,\vec{x}} e^{i \omega t}
 \left\langle 
 \fr12 \left[ \hat H_\rmi{int}(t,\vec{x}), 
 \hat H_\rmi{int}(0,\vec{0}) \right] 
 \right\rangle
 \;.  \la{commu}
\ee
The expectation value
is taken with respect to the density matrix 
of the ``normal'' degrees of freedom which, as alluded 
to above, can be assumed to be thermalized. 

\vspace*{2mm}

\bi 

\centerline{\underline{\hspace*{12cm}}}

\vspace*{2mm}

\item[] {\em Sketch of a proof of \eq\nr{Kubo}}

\vspace*{2mm}

Let us consider the form of \eq\nr{eom} 
in Fourier space, 
with $\varphi \propto e^{-i\omega t + i \vec{p}\cdot\vec{x}}
\;\tilde\varphi$:
\be
 [-\omega^2 + \vec{p}^2 + m_\rmi{eff}^2 - i \omega \Gamma
 + \rmO(\omega^3, \omega \vec{p}^2)] \tilde\varphi 
 = \rmO(\tilde\varphi^2)
 \;. \la{w_eq}
\ee 
We compare the dispersion relation from here
with the ``pole'' appearing in  
the Euclidean propagator of $\tilde\varphi$ after the analytic 
continuation $\omega_n \to -i (\omega + i 0^+)$:
\ba
 && \frac{1}{\omega_n^2 + \vec{p}^2 + \Pi_E 
 }
 \to
 \frac{1}{-\omega^2 -i\omega 0^+ + \vec{p}^2 + 
 \re\Pi_E 
 + i \im \Pi_E 
 }
 \;. 
\ea
Taking $\vec{p}\to\vec{0}$ and  
denoting $m_\rmi{eff}^2 \equiv \re\Pi_E(-i[\omega+i0^+],\vec{0})$, 
we can identify
\be
 \Gamma = - \lim_{\omega\to 0}
 \frac{\im\Pi_E(-i[\omega+i0^+],\vec{0})}{\omega} 
 \;.
\ee
It remains to note that, as can be 
verified with basic path integral techniques in Euclidean spacetime, 
\be
 \Pi_E(\omega_n,\vec{p}) = - 
 \int_{\tau,\vec{x}} e^{i \omega_n \tau + i \vec{p}\cdot\vec{x}}
 \left\langle 
 H_\rmi{int}(\tau,\vec{x}) \, 
 H_\rmi{int}(0,\vec{0})
 \right\rangle
 \;,  \la{PiE}
\ee
where $0 \le \tau \le 1/T$, 
and that its imaginary part yields
\be 
 \im\Pi_E(-i[\omega+i0^+],\vec{0}) = -  \rho(\omega,\vec{0}) 
 \;, \la{rhodef} 
\ee 
with $\rho$ defined through the commutator
in \eq\nr{commu}; \eq\nr{rhodef} will be justified presently. 
This completes the proof of \eq\nr{Kubo}. 

\centerline{\underline{\hspace*{12cm}}}

\ei

\vspace*{5mm}

Before proceeding to prove \eq\nr{rhodef}, 
let us pause to contemplate an important point. 
The oscillation frequency given by \eq\nr{w_eq} does not vanish, but 
is rather given by (for $\vec{p} \to \vec{0}$)
\be
 \omega^2 \approx m_\rmi{eff}^2 \sim m^2 + \frac{T^4}{M^2} 
 \;, \la{meff}
\ee
where we have also included a thermal correction from
the interaction in \eq\nr{L}. Why is it then that $\Gamma$
of \eq\nr{Kubo} is evaluated at vanishing frequency?
The reason is that $\rho(\omega,\vec{0})/\omega$ shows generically
a {\em transport peak} at origin, resembling
a smoothed Dirac $\delta$-function: it has the height $\Gamma$
and a width, $\eta$, determined by the scales of the thermal 
system. 
For Yang-Mills theory, 
parametrically, $\eta \sim g^4 T$~\cite{ms}. 
In the range $m \ll T \ll M$ that 
we are interested in, however, $m_\rmi{eff} \ll T$
according to \eq\nr{meff}. Therefore, 
if $g$ is not exceedingly small, $m_\rmi{eff} \ll \eta \sim g^4 T$, 
and we can well 
approximate $\rho(m_\rmi{eff},\vec{0})/m_\rmi{eff}$ by its limit
$\lim_{\omega\to 0} \rho(\omega,\vec{0})/\omega$, 
simplifying the problem and obtaining in any case 
the {\em largest} (i.e.\ most optimistic) $\Gamma$.

\vspace*{2mm}

\bi 

\centerline{\underline{\hspace*{12cm}}}

\vspace*{2mm}

\item[] {\em Sketch of a proof of \eq\nr{rhodef}}

\vspace*{2mm}

\newcommand{\I}{\int_{-\infty}^{\infty} \!{\rm d}t\,e^{i \omega t}}

Apart from the usual Heisenberg operators 
$
 \hat A(t) = e^{i \hat H t} \hat A(0) e^{-i\hat H t}
$
it is convenient to define ``imaginary-time'' Heisenberg operators as 
$
 \hat A(\tau) \equiv e^{\hat H \tau} \hat A(0) e^{-\hat H \tau}
$, 
where 
$
 0 \le \tau \le \beta
$, 
$\beta \equiv 1/T$.
We furthermore define the correlators  
\ba
  \Pi_{>}(\omega) & \equiv & 
 \I \Bigl\langle \hat A(t) \hat A(0) \Bigl\rangle
 \;,   
 \la{bL}
 \\
  \Pi_{<}(\omega) & \equiv & 
 \I \Bigl\langle \hat A(0)  \hat A(t) \Bigl\rangle
 \;,   
 \la{bS}
 \\
  \rho(\omega) & \equiv & 
 \I \left\langle \fr12 \Bigl[ \hat A(t) , \hat A(0) \Bigr] \right\rangle
 \;, 
 \la{brho}
 \\ 
 G_E(\tau) & \equiv & 
  \Bigl\langle \hat A(\tau)  \hat A(0) \Bigl\rangle
 \;, \\
  \tilde G_E(\omega_n) & \equiv & 
 \int_0^\beta\!{\rm d}\tau \, 
  e^{i \omega_n \tau }
  G_E(\tau)
 \; ; \quad \omega_n \equiv 2 \pi n T  \;, \quad n \in \zz
 \;.
 \la{bE}
\ea 
%
%
Here the expectation value is 
$\langle ... \rangle \equiv \frac{1}{\mathcal{Z}} \Tr 
 \{e^{-\beta \hat H} (...) \}$, and it is
easy to see that $G_E(\beta-\tau)=G_E(\tau)$, i.e.\ 
$G_E(\tau)$ is periodic. Spatial directions have been suppressed
for simplicity but are trivial to incorporate. The correlator $G_E(\tau)$
is theoretically a nice object, because it can be given 
a Euclidean path integral representation and thus a direct
non-perturbative meaning. 

\hspace*{1cm}
Inserting twice $\mathbbm{1} = \sum_n |n\rangle\langle n|$
in \eqs\nr{bL}, \nr{bS}, it is straightforward to show that
$\Pi_{<}(\omega) = e^{-\beta\omega}\Pi_{>}(\omega)$. It then 
follows that 
\be
 \rho(\omega) = \fr12 \left[ \Pi_{>}(\omega) - \Pi_{<}(\omega) \right]
              = \fr12 \left(1 - e^{-\beta\omega} \right) \Pi_{>}(\omega)
 \;, 
\ee 
or, conversely, that 
$
 \Pi_{>}(\omega) = 
 2 \rho(\omega) / (1 - e^{-\beta \omega} )
$.
Making use of the fact that, through their definitions 
as correlators of Heisenberg operators, $G_{>}(t)$ and $G_E(\tau)$ 
are related by the analytic continution $\tau \leftrightarrow it$, 
we finally obtain
\ba 
 \tilde G_E(\omega_n) & = & 
 \int_0^\beta\!{\rm d}\tau \,
 e^{i \omega_n \tau}
 \biggl[ \int_{-\infty}^{\infty} \! \frac{{\rm d} \omega}{2\pi} \, 
  e^{- i \omega t} 
 \Pi_{>}(\omega) \biggr]_{it\to\tau}
 \nn & = & 
 \int_0^\beta \! {\rm d}\tau\, e^{i \omega_n \tau}
 \int_{-\infty}^{\infty} \frac{{\rm d}\omega }{2\pi} e^{-\omega \tau} 
 \Pi_{>}(\omega)
 \nn & = & 
 \int_0^\beta \! {\rm d}\tau\, e^{i \omega_n \tau}
 \int_{-\infty}^{\infty} \frac{{\rm d}\omega }{2\pi} e^{-\omega \tau} 
 \frac{2 \rho(\omega)}{1 - e^{- \beta \omega}} 
 \nn & = & 
 \int_{-\infty}^{\infty} \frac{{\rm d}\omega }{\pi}
 \frac{\rho (\omega)}{1 - e^{-\beta \omega}} 
 \left[ 
   \frac{e^{(i \omega_n - \omega)\tau}}{i \omega_n - \omega}
 \right]^{\beta}_{0}
 \nn & = & 
 \int_{-\infty}^{\infty} \frac{{\rm d}\omega }{\pi}
 \frac{\rho(\omega)}{1 - e^{-\beta \omega}} 
   \frac{e^{- \beta \omega} - 1}{i \omega_n - \omega}
 \nn & = &
 \int_{-\infty}^{\infty} \! \frac{{\rm d} \omega }{\pi} 
 \frac{\rho(\omega)}{\omega - i \omega_n}
 \;. \la{bErhorel} \la{spectral}
\ea
(To be precise, the validity of \eq\nr{spectral} requires the presence
of an ultraviolet regulator for spatial momenta, guaranteeing that both 
$\rho(\omega)$ and $\tilde G_E(\omega_n)$ are cut off at large
values of the argument.)
Equation~\nr{rhodef} now follows by noting that the relation
\be 
 \frac{1}{x \pm i 0^+} = P\left(\frac{1}{x} \right) \mp i \pi \delta(x)
\ee
implies that  
\be
 \im \tilde G_E\left( -i [\omega + i 0^+ ]\right) 
 = 
 \rho(\omega) 
 \;, 
\ee
and by then inserting $-\Pi_E(\omega_n,\vec{p})$ from \eq\nr{PiE} 
for $\tilde G_E(\omega_n)$.

\centerline{\underline{\hspace*{12cm}}}

\ei

\vspace*{5mm}

%
\subsection{Relation of the decay rate $\Gamma$ 
to the bulk viscosity $\zeta$}

Let us now consider the specific case of \eq\nr{L}, {\em viz.}
\be  
 H_\rmi{int} = \frac{1}{M} F^{a\mu\nu}F^a_{\mu\nu}
 \;. \la{Hint}
\ee 
At this point it is good to realize that the structure 
appearing can be recognized as the ``trace anomaly'' of 
pure Yang-Mills theory, 
\be
 \Theta = {T^{\mu}}_\mu \approx 
 -\frac{b_0}{2}\, 
 F^{a\mu\nu}F^a_{\mu\nu} 
 \;, \la{Theta}
\ee
where $b_0$ defines the $\beta$-function related to the running 
coupling, $b_0 = 11\Nc/3(4\pi)^2$. 

\vspace*{2mm}

\bi 

\centerline{\underline{\hspace*{12cm}}}

\vspace*{2mm}

\item[] {\em Sketch of a proof of \eq\nr{Theta}}

\vspace*{2mm}

With the convention 
\be
 S = \int \! {\rm d}t \int \! {\rm d}^{3-2\epsilon}\vec{x} 
 \, \left\{- \frac{1}{4 g_\rmii{B}^2} F^{a\mu\nu}F^a_{\mu\nu} \right\}
\ee
the energy-momentum tensor is 
\be
 T_{\mu\nu} = \frac{1}{g_\rmii{B}^2}
 \left( 
    \fr14 g_{\mu\nu} F^{a\alpha\beta}F^a_{\alpha\beta}
   - {F^{a\alpha}}_{\mu} F^a_{\alpha\nu}
 \right)
 \;. 
\ee
In ${\delta^{\mu}}_{\mu} = 4-2\epsilon $ dimensions its trace is 
\be
 {T^{\mu}}_{\mu} = -\frac{2\epsilon}{4}
 \frac{1}{g_\rmii{B}^2} F^{a\alpha\beta}F^a_{\alpha\beta}
 \;. 
\ee
The inverse of the bare gauge coupling 
($
 g_\rmii{B}^2  =  
 g^2 - \frac{11\Nc}{3\epsilon} \frac{g^4}{(4\pi)^2} + \rmO(g^6)
$)
is \\[-4mm]
\ba
 \frac{1}{g_\rmii{B}^2}  & = &  
 \frac{1}{g^2} + \frac{11\Nc}{3\epsilon} \frac{1}{(4\pi)^2} 
 + \rmO\left(\frac{g^2}{\epsilon}\right)
 \;.
\ea
So, in the limit $\epsilon\to 0$,  
\be
 {T^{\mu}}_{\mu} \approx -\frac{11\Nc}{6} \frac{1}{(4\pi)^2} 
 F^{a\alpha\beta}F^a_{\alpha\beta}
 \;.
\ee

\centerline{\underline{\hspace*{12cm}}}

\ei

\vspace*{5mm}

Returning to \eqs\nr{Hint}, \nr{Theta}, we see that 
$H_\rmi{int} = F^{a\mu\nu}F^a_{\mu\nu} / M \approx -2 \Theta/b_0 M$. 
Therefore \eqs\nr{Kubo}, \nr{commu} advise us to determine
the transport coefficient related to the trace anomaly. However, 
as mentioned after \eq\nr{Tmunu}, the trace of the energy-momentum 
tensor is intimately related to bulk viscosity: in fact the 
latter can be expressed through another Kubo relation, 
\be
 \zeta = \frac{1}{9} \lim_{\omega\to 0} \frac{1}{\omega}
  \int_{t,\vec{x}} e^{i \omega t}
 \left\langle \fr12 
 \left[ \hat \Theta(t,\vec{x}) , \hat \Theta(0,\vec{0}) \right]
 \right\rangle
 \;. \la{zeta_def}
\ee
Furthermore, motivated by the hydrodynamics of heavy ion 
collision experiments,  the weak-coupling
expression for $\zeta$ has been worked out,\cite{adm}  
\be
 \zeta\sim \frac{b_0^2 g^4 T^3}{4 \ln(1/\alpha_s) } 
 \;, 
 \quad \alpha_s \equiv \frac{g^2}{4\pi}
 \;. \la{zeta_res}
\ee
As the appearance of $\ln(1/\alpha_s)$ in the denominator suggests, 
the computation is very non-trivial and necessitates a systematic 
resummation of the perturbative series. 
(Taking the Euclidean formulation as a starting point,
there are also attempts at a non-perturbative 
determination of $\zeta$ in the temperature regime
where the weak-coupling expansion can no longer be justified.\cite{hbm})

In conclusion, combining \eqs\nr{Kubo}, \nr{commu}, 
\nr{zeta_def}, \nr{zeta_res} 
and the relation $H_\rmi{int} \approx -2 \Theta/b_0 M$, 
the vacuum decay rate of the moduli fields, 
$\Gamma \sim {m^3}/{M^2}$, is overtaken at $m \ll T \ll M$ by 
a thermal correction: 
\ba
 \Gamma & = & \lim_{\omega\to 0} \frac{1}{\omega} 
 \int_{t,\vec{x}} e^{i \omega t}
 \left\langle
 \fr12 \left[ \hat H_\rmi{int}(t,\vec{x}) , \hat H_\rmi{int}(0,\vec{0}) 
 \right] \right\rangle
 \nn 
 & \approx & \frac{4\times 9\, \zeta}{b_0^2M^2} \; \sim \; 
 \frac{(12 \pi \alpha_s)^2}{\ln(1/\alpha_s)} \frac{T^3}{M^2} 
 \;\; {\gg} \;\; \frac{m^3}{M^2}
 \;. \la{final_rate}
\ea
In other words, the fact that there is already a plasma present
``facilitates'' the dissipation of the energy in the scalar field
$\varphi$ into that in normal radiation.

%
\subsection{Implications for the moduli problem}

We finally contemplate whether the rate in \eq\nr{final_rate}
could be fast enough to solve the cosmological moduli problem. Since
the rate increases with $T$, we first need to figure out
the initial temperature at which the oscillations start. To do this
properly we would need to rewrite \eq\nr{eom} in an expanding 
background, but the upshot is that oscillations start once their
frequency exceeds the Hubble rate (the inverse of \eq\nr{scale_cos}):
\be
 m_\rmi{eff} \gsim \frac{T^2}{m_\rmi{Planck}} \; \Rightarrow \; 
 T \lsim \sqrt{\rule{0pt}{2ex} m_\rmi{eff}\, m_\rmi{Planck}} \sim \sqrt{m M}
 \;. \la{T_range}
\ee
Then we should integrate the energy loss equation all the way down to 
the temperature $T\sim m$ at which point the vacuum decay takes over. 
We note, however, that within all of this range the decay rate falls
far below the Hubble rate: 
\be
 \frac{(12\pi\alpha_s)^2}{\ln(1/\alpha_s)} \frac{T^3}{M^2}
 \ll \frac{T^2}{M}
\ee
because, according to \eq\nr{T_range},  
$T/M \lsim \sqrt{m/M} \ll 1$. Therefore, the decay is so slow
that it essentially does not have time to take place within the 
lifetime of the Universe; thermal corrections quite probably
{\em cannot} solve the moduli problem.\cite{db}

%
\subsection{Summary}

The purpose of this lecture has been to illustrate various 
generic tools of thermal field theory, as well as the intriguing 
fact that systematic 
heavy ion collision inspired computations may find ``exciting'' 
applications in totally unexpected cosmological contexts. In the 
present example we could learn this way that 
the moduli problem remains a severe constraint even in the 
presence of thermal corrections, a fact
to be taken into account in cosmological model building. 

%
\section{Lecture 3: heavy quark kinetic thermalization}
\la{se:lec3}

The production of a heavy quark and antiquark  
through gluon fusion, which then fly apart in opposite directions, 
is one of the most basic processes in a hadronic collision.
In the heavy ion case, the ``flying apart'' part is non-trivial, 
however, due to the presence of a thermal medium which the heavy
quarks have to surpass. The process is akin to Brownian motion, 
whereby the heavy quarks gradually lose their kinetic energy through 
collisions with the medium particles.\cite{bs,bt} This physics
can be referred to as heavy quark kinetic thermalization, diffusion, 
momentum diffusion, drag, jet quenching, stopping, or energy loss. 
The situation becomes particularly tractable if we focus on 
a ``late stage'' of the process, in which the heavy quarks
are non-relativistic with respect to the heat bath; then it
can be described by Langevin dynamics, with the role of the stochastic
noise being played by the colour-electric Lorentz force. Through 
a QCD generalization of the fluctuation-dissipation theorem
the problem thus boils down to the consideration of the 
2-point temporal correlation function of the colour-electric 
field strength (rendered gauge-invariant by 
time-like Wilson lines).\cite{cs,eucl} The ``transport coefficient''
extracted from this correlator, conventionally referred to 
as $\kappa$, has been
the subject of some recent interest. A leading-order
weak-coupling result\cite{tm} has been supplemented by 
a next-to-leading order correction~\cite{chm}, which has
however been shown to be so large as to question the validity
of the weak-coupling expansion. Numerical simulations have 
been carried out within so-called classical lattice gauge theory, 
confirming the existence of large infrared effects.\cite{mink}
Computations through AdS/CFT techniques for similar processes
in strongly coupled $\mathcal{N}=4$ Super-Yang-Mills theory 
also suggest a much larger $\kappa$ than expected 
in leading-order QCD~\cite{cs,ads}. All of this makes a strong
case for attacking the problem with lattice simulations,  
a challenge that appears technically simpler than 
in the case of many other transport coefficients such as 
viscosities.\cite{rhoE} 
If an answer can be obtained, it can be embedded in 
hydrodynamic simulations of heavy ion collisions (e.g.\ ref.~\cite{ah})
and eventually compared with experimental observations
concerning the spectrum and azimuthal anisotropy
of the decay products from heavy quarks.\cite{exp} 
In cosmology, an analogous kinetic thermalization also 
plays a role, given that many Dark Matter candidates 
kinetically decouple in a non-relativistic regime, 
and their momentum distribution dictates the kind
of structures that can form.\cite{hss}

%
\section{Lecture 4: quarkonium in hot QCD}
\la{se:lec4}

Quarkonium physics in heavy ion collisions is somewhat similar to
heavy quark physics discussed in the previous section, however in 
many ways more complicated: 
initial production is less likely because two heavy quarks
and two heavy antiquarks need to be generated, and furthermore
a quark and antiquark 
have to be in a suitable kinematic range in order to bind together; 
in addition 
propagation through the medium is affected by several processes, 
such as scatterings experienced by the 
quark--antiquark colour dipole, decoherence of 
the quantum-mechanical bound state caused by interactions with the medium, 
as well as Debye screening of the potential that binds the system together.
Recently, some progress has been made in systematizing the study of  
these phenomena, by formulating the problem 
(implicitly or explicitly) 
in an effective field
theory language.~\cite{static} 
The setup makes use of scale separations, 
such as $\alpha_s^2 M \ll \alpha_s M \ll M$ at zero temperature, 
$\alpha_s T \ll \alpha_s^{\scriptscriptstyle 1/2} T \ll T$ 
at finite temperature, and an additional 
$\alpha_s M \ll T \ll \alpha_s^{\scriptscriptstyle 1/2} M$ 
to relate the two.\cite{peskin}
(It remains a challenge to promote 
the setup to the non-perturbative level.)
An example of a generic feature emerging from the effective theory
approach is that a concept of a {\em static potential} can be defined
(as a ``matching coefficient''), 
however it is static in Minkowskian rather than in Euclidean time.
It is also complex unlike at zero temperature, with an imaginary 
part responsible for decoherence or, in frequency space, for the width 
of the quarkonium peak in the corresponding spectral function. 
Thus the static potential is different from that traditionally
extracted from finite-$T$ 
lattice QCD, which is purely real and relies on 
gauge fixing to the Coloumb gauge. Recently, an interesting 
attempt has been launched to extract the proper real-time 
potential from the lattice,\cite{rhs} however further work 
is needed before conclusions can be drawn. Another insight 
is that purely perturbative studies may converge much better than 
in general (cf.\ the previous section),\cite{singlet} 
because only short-distance 
physics plays a role for heavy quarkonium.
Ultimately, one of the goals of these efforts is to determine 
the spectrum of the dilepton pairs produced from thermal 
quarkonium decays\cite{NLO} (given that quarks disappear we may refer 
to this as an inelastic process). A cosmological analogue 
may exist in near-threshold two-particle Dark Matter production 
or decay, relevant for some of the most popular scenarios.\cite{gg,md}

%
\section*{Acknowledgements}

I am grateful to Dietrich B\"odeker for interesting discussions.
The work was supported in part by the Yukawa International Program 
for Quark-Hadron Sciences at Yukawa Institute for 
Theoretical Physics, Kyoto University, Japan; 
I wish to thank the Organizers of the Program for the 
invitation and for the kind hospitality. 

%

\end{document}